\documentclass{webofc}
\usepackage[varg]{txfonts}   
\usepackage{amsmath}
\begin{document}

\title{A new approach to stochastic relativistic fluid dynamics from information flow}
\date{\today}

\author{\firstname{Nicki} \lastname{Mullins}\inst{1}\fnsep\thanks{\email{nickim2@illinois.edu}} \and
        \firstname{Mauricio} \lastname{Hippert}\inst{1}\fnsep\thanks{\email{hippert@illinois.edu}} \and
        \firstname{Lorenzo} \lastname{Gavassino}\inst{2}\fnsep\thanks{\email{lorenzo.gavassino@vanderbilt.edu}} \and \firstname{Jorge} \lastname{Noronha}\inst{1}\fnsep\thanks{\email{jn0508@illinois.edu}}
}

\institute{Illinois Center for Advanced Studies of the Universe \& Department of Physics, University of Illinois Urbana-Champaign, Urbana, IL 61801, USA 
\and
           Department of Mathematics, Vanderbilt University, Nashville, TN, USA
          }

\abstract{We present a new general formalism for introducing thermal fluctuations in relativistic hydrodynamics, which incorporates recent developments on the causality and stability of relativistic hydrodynamic theories. Our approach is based on the information current, which measures the net amount of information carried by perturbations around equilibrium in a relativistic many-body system. The resulting noise correlators are guaranteed to be observer-independent for thermodynamically stable models. We obtain an effective action within our formalism and discuss its properties.}   

\maketitle

\section{Introduction}

The matter formed during high-energy heavy-ion collisions is modeled as an expanding viscous relativistic fluid known as the quark-gluon plasma (QGP). While this modeling has had great success describing data, it is known that any dissipative system near equilibrium exhibits spontaneous thermal fluctuations, which is the physics content behind the fluctuation-dissipation theorem \cite{Callen:1951vq, Kubo:1957mj}. Therefore,  self-consistent hydrodynamic simulations of the QGP should also investigate these stochastic fluctuations. 


In this proceedings, we report on a new approach for including stochastic fluctuations in relativistic systems developed in  \cite{Mullins:2023tjg,Mullins:2023ott}. This formalism is guaranteed to be causal and stable against fluctuations around the equilibrium state in a Lorentz-invariant manner. Our construction relies on the information current \cite{Gavassino:2021kjm}, a quantity that measures how much information is contained in a given perturbation around the equilibrium state. This information current can be used to obtain the equilibrium probability distribution for the system to be in a given state. We also construct an effective action for the fluctuating hydrodynamic system and show how the noise can be obtained from a symmetry of this action. 

\section{The information current}

The free energy describes the thermal state of a system at a given time. To define this in relativity, we must foliate the spacetime into a set of spacelike hypersurfaces $\Sigma_{\tau}$ with past-directed timelike unit normal vectors $n^{\mu}$. The free energy depends on the thermodynamic variables and how we choose to foliate spacetime. On a given hypersurface, the free energy variation from equilibrium, $\delta \Omega$, is given by 
\begin{equation}
    \delta \Omega = T \int_{\Sigma_{\tau}} d\Sigma n_{\mu} \left( - \delta s^{\mu} - \alpha_I \delta J^{\mu I} \right) ,
\end{equation}
where $\delta s^{\mu}$ is the variation of the entropy current from equilibrium, $\alpha_I$ are the thermodynamic conjugates to the conserved quantities evaluated at equilibrium, and $\delta J^{\mu I}$ are the variations of the conserved currents from equilibrium. The quantity in parentheses is the information current \cite{Gavassino:2021kjm}
\begin{equation}
    E^{\mu} = - \delta s^{\mu} - \alpha_I \delta J^{\mu I} .
\end{equation}
The probability distribution for the system to be in a given macroscopic state around equilibrium is then given by 
\begin{equation} \label{Eq:Equilibrium_dist}
    p[\delta \phi^A] \sim e^{-\delta \Omega/T}= e^{- \int d\Sigma n_{\mu} E^{\mu}} ,
\end{equation}
where $\delta \phi^A$ is a vector containing the hydrodynamic fields of the system. Note that this probability distribution seemingly depends on the choice of foliation. However, physical results should be observer-independent, so ensuring that this dependence on the foliation drops out when calculating the correlation functions for observable quantities is important. 

While the utility of the information current should already be apparent through its connection to thermodynamics, it also has deep connections to causality and stability. In particular, it has been shown in \cite{Gavassino:2021kjm} that a relativistic system will be (linearly) causal and stable against fluctuations if:
\begin{enumerate}
    \item $n_{\mu} E^{\mu} \geq 0$ for any past-directed timelike $n^{\mu}$.  
    \item The bound $n_{\mu} E^{\mu} = 0$ is saturated if and only if $\delta \phi^A = 0$, hence equilibrium is unique. 
    \item Finally, the second law of thermodynamics, $\partial_{\mu} E^{\mu} \leq 0$ must hold. 
\end{enumerate}
As long as we ensure that the information current satisfies these conditions, our fluctuating theory will be linearly causal and stable. To take advantage of these properties, we construct a theory of fluctuations directly from the information current.

\section{Fluctuations from information flow}

The linearized dynamics of a relativistic hydrodynamic system can be extracted from the fact that $\partial_{\mu} E^{\mu} = -\sigma$, where $\sigma$ is the entropy production. Using this, we can express the equations of motion in terms of the hydrodynamic variables as 
\begin{equation} \label{Eq:EoM}
    \left( E_{AB}^{\mu} \partial_{\mu} + \sigma_{AB} \right) \delta \phi^B = \xi_A ,
\end{equation}
where $E^{\mu} = \delta \phi^A E_{AB}^{\mu} \delta \phi^B / 2$, $\sigma = \delta \phi^A \sigma_{AB} \delta \phi^B$, and $\xi_A$ is a Gaussian stochastic vector with zero mean. This approach is particularly well-suited for Israel-Stewart-like hydrodynamic models \cite{ISRAEL1979341} as these are constructed from the entropy current, conserved quantities, and entropy production; the same ingredients are used here. Using this equation of motion, we can write the momentum space correlation function of the hydrodynamic variables as 
\begin{equation}
    \langle \delta \phi^A \delta \phi^B \rangle = \left( i E^{\mu AC} k_{\mu} + \sigma^{AC} \right)^{-1} \langle \xi_C \xi_D \rangle \left( i E^{\mu DB} k_{\mu} + \sigma^{DB} \right)^{-1 \dagger} .
\end{equation}
These correlation functions should match those obtained from the equilibrium probability distribution of Eq.\ \eqref{Eq:Equilibrium_dist}, which can be used to determine the form of the noise correlator $\langle \xi_A(x) \xi_B(x') \rangle$. In particular, it is found that
\begin{equation} \label{Eq:noise_correlators}
    \langle \xi_A(x) \xi_B(x') \rangle = 2 \sigma_{AB} \delta^{(4)}(x-x') ,
\end{equation}
for thermodynamically stable systems. The full details of this derivation are provided in \cite{Mullins:2023tjg}, as well as proof that the corresponding noise correlators do not depend on the choice of foliation. 

\subsection{Fluctuations in relativistic diffusion}

We can apply this to the Israel-Stewart theory of a conserved current in the Landau hydrodynamic frame. Such a system is defined by the conserved current $J^{\mu} = n u^{\mu} + \mathcal{J}^{\mu}$, 
where $n$ is some density, $u^{\mu}$ is the fluid velocity (with $u_{\mu} u^{\mu} = -1$), and $\mathcal{J}^{\mu}$ is the dissipative part of the conserved current (where $\mathcal{J}^{\mu}u_\mu=0$). The corresponding entropy current is given by 
\begin{equation}
    s^{\nu} = s u^{\nu} - \frac{\mu}{T} \mathcal{J}^{\nu} - \frac{\beta_J u^{\nu}}{2T} \mathcal{J}_{\lambda} \mathcal{J}^{\lambda} ,
\end{equation}
where $s$ is the equilibrium entropy density, $\mu$ is the chemical potential associated with the density $n$, and $\beta_J$ is some new transport coefficient. The entropy production should be positive definite, so we take it to be a quadratic form $\sigma = \frac{1}{\kappa T} \mathcal{J}_{\lambda} \mathcal{J}^{\lambda}$, where $\kappa$ is the charge conductivity. The information current is then given by 
\begin{equation}
\begin{split}
    E^{\mu} & = \frac{u^{\mu}}{2\chi T} \delta n^2 + \frac{1}{\chi T} \delta n\, \delta \mathcal{J}^{\mu} + \frac{\beta_J u^{\mu}}{2T} \delta J_{\lambda} \delta J^{\lambda} ,
\end{split}
\end{equation}
where $\chi = \partial n / \partial \mu$. It can be verified that the equations of motion from Eq.\ \eqref{Eq:EoM} are the linearized versions of the conservation law and the Israel-Stewart relaxation equation for $\mathcal{J}^{\mu}$. From Eq.\ \eqref{Eq:noise_correlators}, it follows that the conservation law does not fluctuate, while the relaxation equation has a stochastic source, $\xi^{\mu}_{\perp}$, with noise correlator 
\begin{equation}
    \langle \xi^{\mu}_{\perp}(x) \xi^{\nu}_{\perp}(x') \rangle = \frac{1}{\kappa T} \Delta^{\mu\nu} \delta^{(4)}(x-x') .
\end{equation}
Here, $\Delta^{\mu\nu} = g^{\mu\nu} + u^{\mu} u^{\nu}$ is the projector orthogonal to $u^{\mu}$. From this, the symmetrized momentum space correlator for the density $n$ can be easily obtained. One finds
\begin{equation}
    \langle \delta n \delta n \rangle = \frac{\kappa \chi^2 \omega k^2}{|\kappa k^2 - i \chi \omega - \kappa \chi \beta_J \omega^2|^2} ,
\end{equation}
which reduces to the standard first-order result in the suitable limit \cite{Mullins:2023tjg}.

\subsection{Actions for fluctuating hydrodynamics}

Stochastic partial differential equations can be written as a path integral over the dynamical variables and some set of auxiliary variables \cite{Martin:1973zz}. For systems obeying Eq.\ \eqref{Eq:EoM}, we find that the effective Lagrangian is 
\begin{equation} \label{Eq:effective_action}
    \mathcal{L}_{\mathrm{EFT}} = -\delta \bar{\phi}^A \left( E_{AB}^{\mu} \partial_{\mu} + \sigma_{AB} \right) \delta \phi^B + i \delta \bar{\phi}^A \sigma_{AB} \delta \bar{\phi}^B .
\end{equation}
Here, $\delta \bar{\phi}^A$ are the auxiliary variables. The first term corresponds to the non-fluctuating equation of motion, while the second term gives the fluctuations.

Above, we have inserted the fluctuation-dissipation relation of Eq.\ \eqref{Eq:noise_correlators}, but detailed balance provides a means to implement the fluctuation-dissipation theorem at the level of the action. Let $\Theta$ denote a transformation under time reversal and parity. Then, employing a similar derivation to that used in \cite{Guo:2022ixk}, we find that the action should transform as $\mathcal{L}_{\Theta} = \mathcal{L} + i \delta \phi^A E_{AB}^{\mu} \partial_{\mu} \delta \phi^B$
under a transformation involving time reversal and parity. The appropriate symmetry for a system with an effective action of the form Eq.\ \eqref{Eq:effective_action} is
\begin{equation}
    \delta \phi^A \rightarrow \Theta \delta \phi^A , \:\: \delta \bar{\phi}^A \rightarrow -\Theta \delta \bar{\phi}^A - i \Theta \delta \phi^A .
\end{equation}
This symmetry can be used to determine the noise correlators, implementing the fluctuation-dissipation theorem. In principle, nothing about the derivation used to obtain this result required the equations of motion to be linear. Hence, we expect this property to be valid in nonlinear systems \cite{Huang:2023eyz}. Another approach for constructing effective actions for Israel-Stewart theories was recently presented in \cite{Jain:2023obu}.

\section{Conclusions}

In \cite{Mullins:2023tjg}, a formalism for stochastic fluctuations was constructed using the information current. This approach incorporates causality and stability conditions built into the information current to ensure that noise correlators are independent of the choice of spacetime foliation. This new framework is easily applied to linearized Israel-Stewart-like theories, as demonstrated by the example involving a fluctuating conserved current. Using the Martin-Siggia-Rose approach, it is possible to derive an effective action for any fluctuating system with an information current and non-negative entropy production \cite{Mullins:2023ott}. Demanding that the path integral be consistent with the principle of detailed balance, we found a symmetry of the effective action involving time reversal and parity that can implement the fluctuation-dissipation theorem. This symmetry is derived without any assumptions regarding linearity, indicating that it should be possible to generalize the results presented here to nonlinear systems. 

\section*{Acknowledgements}

NM and JN are partially supported by the U.S. Department of Energy, Office of Science, Office for Nuclear Physics
under Award No. DE-SC0021301 and DE-SC002386. MH and JN were partially funded by the National Science Foundation within the framework of the MUSES Collaboration under grant number OAC-2103680. LG is partially supported by Vanderbilt's Seeding Success Grant.

\bibliography{references}

\end{document}